\documentclass[aps,prx,twocolumn,superscriptaddress,amsmath,amssymb,nofootinbib,10pt]{revtex4-1}
\usepackage{graphicx}
\usepackage{amsmath,amssymb,bm} 
\usepackage{ulem}
\usepackage{soul}
\usepackage{cancel}
\usepackage{upgreek}
\usepackage{txfonts}
\usepackage[colorlinks, linkcolor=blue]{hyperref}
\usepackage{color} 
\usepackage[papersize={8.5in,11in}]{geometry}
\usepackage{verbatim}
\usepackage{esint}
\hypersetup{
    unicode=false,          
    pdftoolbar=true,        
    pdfmenubar=true,        
    pdffitwindow=false,     
    pdfstartview={FitH},    
    pdftitle={},    
    pdfauthor={},     
    pdfsubject={},   
    pdfcreator={},   
    pdfproducer={}, 
    pdfkeywords={} {} {}, 
    pdfnewwindow=true,      
    colorlinks=true,       
    linkcolor=magenta, 
    citecolor=blue,        
    filecolor=magenta,      
    urlcolor=blue           
} 

\renewcommand{\vec}{\bf}
\def \k {{\vec k}}

\def \a {{\vec a}}
\def \b {{\vec b}}

\def \ve {\varepsilon}
\def \r {{\vec r}}

\def \q {{\vec q}}
\def \d{\partial}
\def \Q{{\vec Q}}

\def \ve {\varepsilon}

\def \S {\vec{S}}

\def \D{\Delta}
\def \A{{\vec A}}

\def \L{{\cal{L}}}

\def \beq {\begin{eqnarray}}
\def \eeq {\end{eqnarray}}
\def \tn {\textnormal}

\begin{document}

\title{Mixed-valence insulators with neutral Fermi surfaces}
\author{Debanjan Chowdhury} 
\thanks{Corresponding author: debch@mit.edu}
\affiliation{Department of Physics, Massachusetts Institute of Technology, Cambridge, Massachusetts
02139, USA.}
\author{Inti Sodemann}
\affiliation{Department of Physics, Massachusetts Institute of Technology, Cambridge, Massachusetts
02139, USA.}
\affiliation{Max-Planck Institute for the Physics of Complex Systems, Dresden, 01187, Germany.}
\author{T. Senthil}
\affiliation{Department of Physics, Massachusetts Institute of Technology, Cambridge, Massachusetts
02139, USA.}
\begin{abstract}
{\bf Abstract-} Samarium hexaboride is a classic three-dimensional mixed valence system with a high-temperature metallic phase that evolves into a paramagnetic charge insulator below 40 kelvin. A number of recent experiments have suggested the possibility that the low-temperature insulating bulk hosts electrically neutral gapless fermionic excitations. Here we show that a possible ground state of strongly correlated mixed valence insulators---composite exciton Fermi liquid--- hosts a three dimensional Fermi surface of a neutral fermion, that we name the  ``composite exciton". We describe the mechanism responsible for the formation of such excitons, discuss the phenomenology of the composite exciton Fermi liquids and make comparison to experiments in SmB$_6$.  
\end{abstract}

\maketitle
\begin{center}
{\bf Introduction}
\end{center} 
Electronic solids where the valence of one of the constituent elements is non-integral show a number of fascinating properties \cite{MottMV,CMV76} arising from the Coulomb interaction between electrons. 
Of interest to us in this paper is a class of mixed-valence (MV) systems, a classic example being SmB$_6$ \cite{Geballe69,Geballe71}, where a high temperature metallic state evolves into an insulator at low temperatures. Attention has been refocused on this material in recent years following the proposal \cite{Coleman1} that it may be an interaction-driven topological insulator (TI) \cite{Kane10,Zhang11}. 
There is compelling evidence now for metallic surface states in this material (of possibly topological origin) despite an electrically insulating bulk at low temperatures from predominantly transport \cite{Batlogg1,paglione1,fisk1,fisk2,mcqueen1,balakrishnan} and other measurements \cite{TKIrev}. A different fascinating aspect of a number of MV insulators, including  SmB$_6$ \cite{Geballe69,Geballe71},  are various thermodynamic and transport anomalies at low temperatures, apparently at odds with an insulating behavior in the bulk. Traditionally, these anomalies have often been attributed to the presence of in-gap states.  An interesting development was the observation of quantum oscillations (QO) in magnetization, first reported in SmB$_6$  by Li et al. \cite{Li14} and interpreted as additional evidence for the two-dimensional metallic surface states. 

 However, subsequent measurements of QO in magnetization in SmB$_6$ by Tan et al. \cite{SS15} observed frequencies corresponding to almost half of the bulk Brillouin zone. Tan et al. \cite{SS15} found that the frequencies, the cyclotron mass and the amplitude of the oscillations are quite similar to the measured quantum-oscillations in the other metallic hexaborides $R$B$_6$ ($R\equiv$ La, Pr, Ce) \cite{LB6,LB62,PB6,CB6}. Moreover the measured density of states from the low-temperature specific heat is in good agreement with the value obtained from quantum oscillations \cite{SSnew}. Based on these observations Ref. \onlinecite{SS15} raised the surprising possibility that the quantum oscillations are a property of the electrically insulating bulk. They also suggested that the oscillations originate from the same in-gap states responsible for the low temperature anomalies which have since been re-examined closely. However it has also been argued more recently \cite{Den16} that some of the same QO results can be explained using a purely two-dimensional model of the metallic surface states. The low temperature anomalies include a finite linear specific heat coefficient \cite{flachbart,SS15,thompson} and bulk optical conductivity below the charge-gap \cite{Armitage16}. Furthermore, a field-induced thermal conductivity proportional to the temperature has been reported in some samples \cite{SSnew}  (though this feature does not seem to be present universally \cite{Li16,LT17}). Taken together these measurements suggest the presence of a Fermi surface of electrically neutral fermions in the bulk that nevertheless couple to the external magnetic, but not to weak DC electric-fields. 

Inspired by the current baffling experimental situation, we are led to a number of theoretical questions. Can MV insulators host Fermi surfaces of neutral fermionic quasiparticles? If so, what is the origin of the neutral (fermionic) excitation and what constrains the volume of the Fermi surface? What are the thermodynamic and transport signatures of phases with such neutral fermionic excitations? Can Fermi surfaces of neutral fermions, that do not couple directly to the external magnetic-field, give rise to quantum oscillations? In a separate development, it has been pointed out \cite{CooperMB,FWMB} that under certain conditions, even band-insulators with gaps smaller than the cyclotron energy can exhibit quantum oscillations. 

In recent years, a number of triangular lattice organic materials close to the Mott transition have been shown to act as charge-insulators but thermal metals \cite{Shimizu03,Yamashita08,Yamashita10,Yamashita11}, where the electron appears to have splintered apart into fractionalized excitations (``partons") \cite{PAL05,OM05}; while the charge degree of freedom can remain gapped, the spinful, neutral spinon can form a Fermi surface. The possibility of observing quantum oscillations for such neutral spinon Fermi surfaces has been addressed previously by Motrunich \cite{OM06}. However, strongly correlated mixed-valent insulators are far from being a Mott insulator, thereby requiring a different microscopic mechanism to stabilize such a neutral Fermi surface. 

Here we show that in the limit of strong Coulomb interactions in a mixed-valence insulator, there is a well defined mechanism for the formation of an electrically neutral fermionic quasiparticle---dubbed the fermionic composite exciton (ce)---that  can form a Fermi surface; the resulting phase - the composite exciton Fermi liquid (CEFL) -  is electrically insulating but will have a neutral fermi surface. We show that the CEFL shares a number of features with the observed phenomenology in SmB$_6$. We also note that while the present work is motivated by the recent experiments in SmB$_6$, it is potentially relevant to other mixed-valence insulators \cite{CMV76,TKIrev}, such as SmS under pressure, YbB$_{12}$ etc. 

\begin{center}
{\bf Results}
\end{center}

{\bf Electronic structure-} The electronic configuration of Sm is [Xe]$4f^6 5d^0 6s^2$. In SmB$_6$, the valence of Sm is known to fluctuate between Sm$^{2+}$ and Sm$^{3+}$ with an average valence of approximately $\sim 2.6$ \cite{valence,valence2}. There is strong spin-orbit coupling in this material and the six-fold degeneracy of the $J=\frac{5}{2}$ orbital is lifted due to crystal field splitting, giving rise to a quartet ($\Gamma_8$) and a doublet ($\Gamma_7$). The five-fold degenerate $d-$orbitals split up into a doublet ($e_g$) and a triplet ($t_{2g}$). The ground state of Sm in SmB$_6$ is in a coherent superposition of $5d^1~ (e_g) + 4f^5~ (\Gamma_8) \leftrightharpoons 4f^6$. In contrast, the ground state of La in metallic LaB$_6$ has an electronic configuration of [Xe]$5d^1 6s^2$ and there are no $f-$electrons. 

Band-structures for the surface as well as the insulating bulk \cite{Dai13} have been modeled using multi-orbital tight-binding models \cite{coleman2,MV15}, but we focus here on the simplest two-band model for a mixed-valence compound in three dimensions \cite{CMV76} to illustrate the key ideas. In particular, we will restrict ourselves to the situation where both the $d$ and $f$ orbitals are treated as doublets instead of quartets. 

{\bf Model-} We start with a band of $d-$conduction electrons, where $d_{\r\sigma}$ is the annihilation operator for a $d-$electron at site $\r$ with spin $\sigma$, and a heavy band of $f$ electrons, where $f_{\r\alpha}$ is the annihilation operator for an $f-$electron at site $\r$ and crystal-field multiplet index $\alpha$ (both $\sigma, \alpha = \uparrow, \downarrow$ and we drop the distinction between the two from now on). As discussed above, it is appropriate to consider a model where the $f-$valence fluctuates between $n^f=1$ and $n^f=2$. With respect to the $n^f=2$ state, the above configurations can be interpreted as an empty state and a state with one hole respectively. We therefore carry out the following particle-hole (PH) transformation $f_\alpha\rightarrow \ve_{\alpha\beta} f_\beta^\dagger = \tilde{f}_\alpha$
where $\ve_{\alpha\beta}$ is the fully antisymmetric tensor and we have introduced $\tilde{f}$ as the $f-$hole. The standard periodic Anderson Hamiltonian \cite{VY}, but now written in terms of the $\tilde{f}-$hole is given by, 
\beq
H &=& \sum_{\r\r',\alpha} (-t^d_{\r\r'} - \mu_d ~\delta_{\r\r'}) d_{\r\alpha}^\dagger d_{\r'\alpha} - \sum_{\r\r',\alpha}t_{\r\r'}^f \tilde{f}_{\r\alpha}^\dagger \tilde{f}_{\r'\alpha}\nonumber\\
&&+ \sum_{\r, \r'} \bigg[\ve_{\beta\gamma} V_{\alpha\beta}(\r-\r') d^\dagger_{\r\alpha} \tilde{f}^\dagger_{\r'\gamma} + \tn{H.c.}\bigg] - U_{df}\sum_\r n^{\tilde{f}}_\r n^d_\r \nonumber\\
&&+ U_{ff} \sum_\r n^{\tilde{f}}_\r(n^{\tilde{f}}_\r-1),
\label{MVham}
\eeq
where $n^{\tilde{f}}_\r = \sum_\alpha \tilde{f}_{\r\alpha}^\dagger \tilde{f}_{\r\alpha} = 2 - n^f_\r$, with $n_\r^f = \sum_\alpha f^\dagger_{\r\alpha} f_{\r\alpha}$ and $n_\r^d = \sum_\alpha d^\dagger_{\r\alpha} d_{\r\alpha}$. $U_{df}$ is a repulsive density-density interaction between the $f$ and $d-$electrons (or equivalently, it represents an attractive interaction between the $\tilde{f}-$hole and the $d-$electron) and $U_{ff}$ represents a large on-site Coulomb repulsion between the $f-$electrons. The hoppings for the $d-$electron ($\tilde{f}-$hole) are given by $t^d_{\r\r'}$ ($t^f_{\r\r'}$), with $|t^d|\gg |t^f|$ and $\mu_d$ represents the chemical-potential for $d-$electrons.The hybridization, $V_{\alpha\beta}$, between the $d$ and $f$ electrons,  has odd parity $V_{\alpha\beta}(-\k) = -V_{\alpha\beta}(\k)$.

\begin{figure}
\begin{center}
\includegraphics[width=1.0\columnwidth]{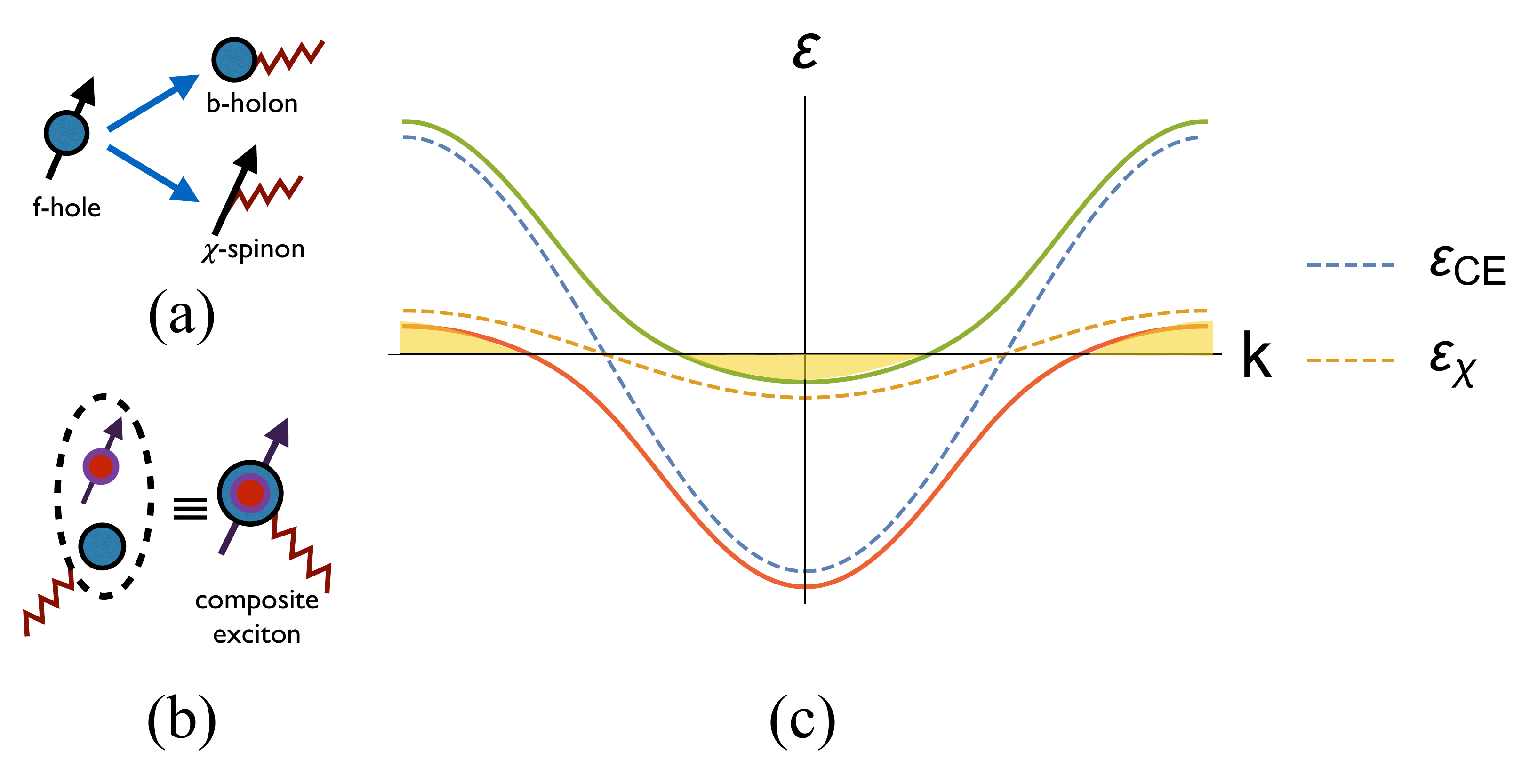}
\end{center}
\caption{{\bf Route to composite exciton Fermi liquids.} {\bf (a)} Slave-boson representation for the $\tilde{f}-$hole in terms of a holon (blue circle) and spinon (black arrow), coupled mutually to $\a$ (zigzag line).   {\bf (b)} Strong binding of the conduction $d-$electron (red circle with arrow) to the holon leads to formation of a fermionic composite exciton coupled to the same $\a$. {\bf (c)} Two-band model when the $f-$valence fluctuates between $n^f = 1$ and $2$. The composite exciton dispersion $\ve_{\tn{CE}}$ (blue dashed line) and a narrow spinon dispersion $\ve_\chi$ (orange dashed line) shown for the gauge-invariant combination $\zeta = t_{\tn{CE}}/t_\chi>0$. The hybridization between the two gives rise to two bands (orange and green solid lines) and as a result of the filling leads to a semi-metallic state (yellow shaded regions), where the volumes of the two pockets are equal. For $\zeta<0$, the resulting state would be an insulator.}
\label{disp}
\end{figure}

We are interested in the limit of $U_{ff}\rightarrow\infty$, and $U_{df}$ large but finite. We use here a slightly different variant of the standard slave-boson representation \cite{colemanslave},
\beq
\tilde{f}_{\r\alpha} = b_\r \chi_{\r\alpha},
\label{sb}
\eeq
 where we have fractionalized the $\tilde{f}-$hole into (i) a spinless boson (``holon"), $b$, that carries the physical, negative $(-1)$ electromagnetic charge (i.e. opposite to electron charge) under the external gauge-field, $A_\mu$, and, (ii) a neutral fermion (``spinon"), $\chi_\alpha$, that carries the spin ($\alpha$); see Fig.\ref{disp}a. There is a redundancy associated with the above parametrization $\chi\rightarrow\chi_\alpha e^{-i\theta},~b\rightarrow b e^{i\theta}$ which leaves $f_\alpha$ invariant. We therefore assume that the holon (spinon) carries a unit positive (negative) charge under an emergent $U(1)$ gauge-field $a_\mu = (a_0,\a)$. We are interested in describing phases with a charge-gap (i.e. insulators) where the holon remains uncondensed, $\langle b\rangle = 0$ and where the Fermi surface of the $d-$electrons is absent. The above definition in terms of the partons is to be supplemented with a gauge-constraint, that ensures restriction to gauge-invariant states in the Hilbert space, of the form $b^\dagger_\r b_\r = \chi_{\r\alpha}^\dagger \chi_{\r\alpha}$.  We impose an additional hard-core constraint on the bosons, i.e. $b_\r^\dagger b_\r \leq 1$, which ensures no double occupancy of the $\tilde{f}-$hole; the total density of doped holes is then $\sum_\r b^\dagger_\r b_\r = \sum_\r \tilde{f}^\dagger_{\r\alpha} \tilde{f}_{\r\alpha}$. (See the Methods section for a comparison to the standard slave-boson representation.)  
 
{\bf Composite excitons-} The global requirement for obtaining a mixed-valence insulator, that is also consistent with the known electronic count in SmB$_6$ is $ \sum_\r \tilde{f}^\dagger_{\r\alpha} \tilde{f}_{\r\alpha} = \sum_\r d_{\r\sigma}^\dagger d_{\r\sigma}$ (equivalently, $\sum_\r [d_{\r\sigma}^\dagger d_{\r\sigma} + f_{\r\alpha}^\dagger f_{\r\alpha}] = 2$), which when combined with the above constraints automatically implies $n^b = n^d$. As a result of the attractive interaction (Eq.\ref{MVham}) between the $\tilde{f}-$holes and $d-$electrons ($ U_{df}>0$), there is now an attractive interaction between the holons and the conduction electrons. For sufficiently strong attraction, it is therefore possible to form bound states of the conduction electrons and the holons to form a neutral fermionic composite exciton (fCE),
\beq
\psi_{\k\alpha} \equiv b~ d_{\k\alpha},~~\psi_{\k\alpha}^\dagger \equiv b^*~ d_{\k\alpha}^\dagger.
\eeq
The above quasiparticle is electrically neutral but is charged under the internal $U(1)$ gauge field associated with the slave boson construction (see fig.\ref{disp}a); at a finite density it can form a Fermi surface that is minimally coupled to the emergent gauge-field $a_\mu$. Note that in our specific example, as a result of the hard-core constraint, the number of bosons are guaranteed to be equal to the number of conduction electrons and therefore, the number of fCE is identical to the number of conduction electrons, i.e. $n^\psi = n^d$. The volume of the Fermi surface of the $\psi$ fermions will then be identical to the volume of the original conduction ($d-$)electron Fermi surface. 

The effective Hamiltonian that describes the low energy physics, after the conduction electrons have formed bound states with the holons, can be expressed as,
\beq
H' &=& \sum_{\k,\alpha} \ve_{\tn{CE}} ~\psi_{\k\alpha}^\dagger \psi_{\k\alpha} + \sum_{\k,\alpha} \ve_{\chi,\k}~\chi_{\k\alpha}^\dagger \chi_{\k\alpha}\nonumber\\
&+& \sum_{\r\r',\alpha\beta} \bigg[\ve_{\beta\gamma} V_{\alpha\beta}(\r-\r') \psi^\dagger_{\r\alpha} \chi^\dagger_{\r'\gamma} + \tn{H.c.}\bigg] + ...,
\label{MV2n}
\eeq
where $\ve_{\tn{CE}}$ is the fCE dispersion (see Methods section for an estimate of the nearest neighbor fCE hopping) and $\ve_{\chi,\k}$ is the spinon dispersion. Note that, by construction, the holon is gapped. On the other hand as a result of the complete binding of all the $d-$electrons to form fCE, the charged $d-$excitations are also gapped. The ellipses denote various allowed terms; one such term (among others) is the exchange interaction between the $f$ moments, 
\beq
H_{\tn{ex}} = J_{\tn H}\sum_{\langle\r,\r'\rangle} \S_\r\cdot\S_{\r'},
\eeq
which also modifies the dispersion for the spinon bands, with the hopping $t_\chi$ set by $t^f$, $J_{\tn H}$ and the holon hopping (see supplementary note 1).

For a finite $V$, the fCE band hybridizes with the spinon band to yield renormalized bands as shown in fig. \ref{disp}(b) (see Methods section). It is convenient to carry out a PH transformation on $\chi_{\r\alpha}\rightarrow\tilde\chi_{\r\alpha} \equiv \ve_{\alpha\beta} \chi_{\r\beta}^\dagger$. Then,
\beq
\sum_\r n^\psi_\r = \sum_\r n_\r^\chi = \sum_\r (2 - n_\r^{\tilde{\chi}}).
\eeq
 A finite $t_\chi$ is necessary to get crossings at the fermi-level; one then obtains an electrically neutral semi-``metal" with `particle' and `hole' pockets with equal volume.  The Fermi surfaces thus obtained have both fCE and spinon character; from now on we do not distinguish between the two. Note that the hopping amplitudes for the fCE and spinon are not individually gauge-invariant, unlike the gauge-invariant ratio $\zeta = t_{\tn{CE}}/t_\chi$. Which sign of $\zeta$ is preferred depends on various microscopic details; if $\zeta<0~(\zeta>0)$ the ground state will in fact be an insulator (semi-metal) of fCE and spinons.

\begin{figure}
\begin{center}
\includegraphics[width=0.8\columnwidth]{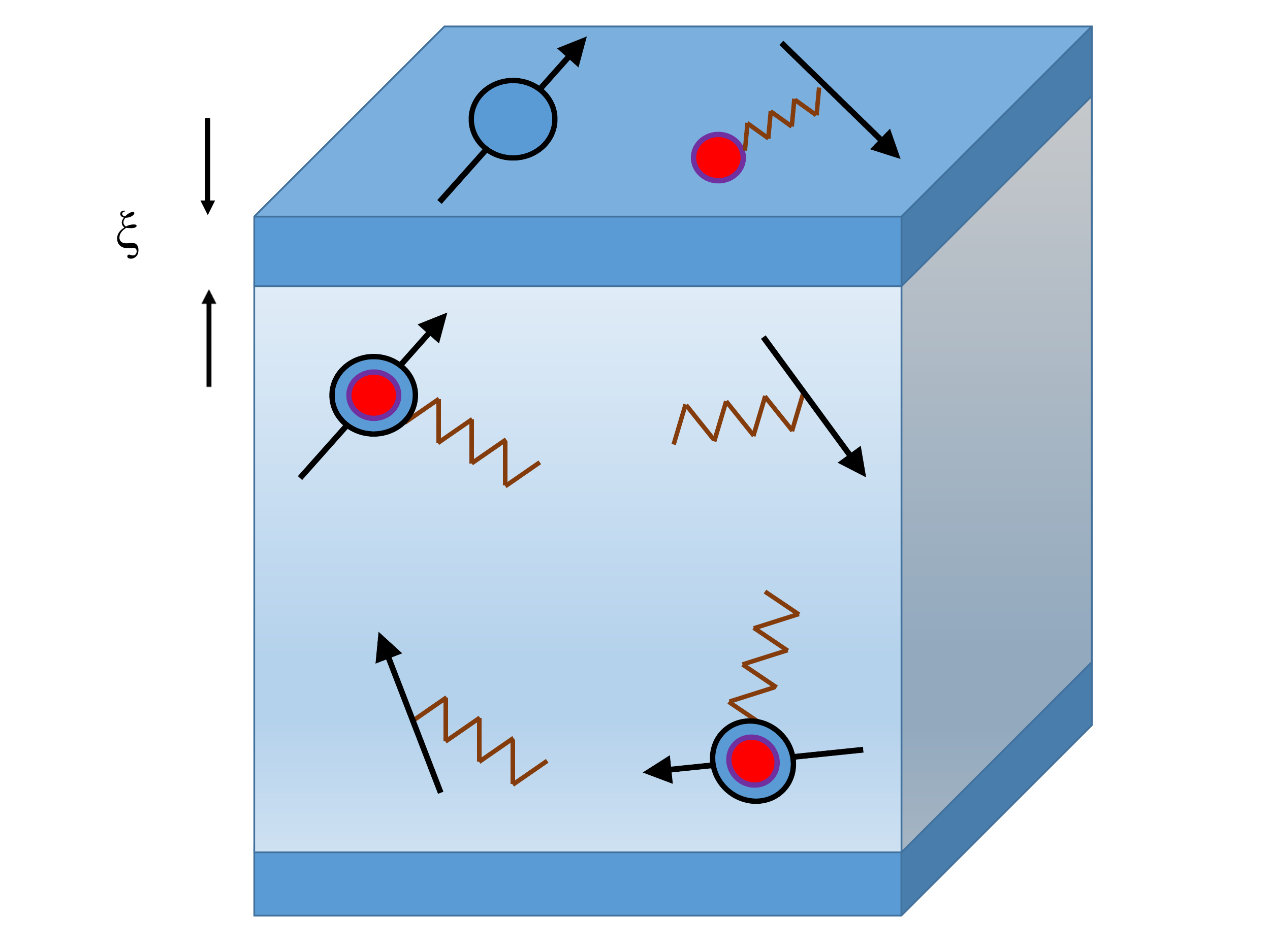}
\end{center}
\caption{{\bf Metallic surface with insulating CEFL bulk.} The bulk realizes a CEFL, a compensated semi-metal with particle and hole-like pockets (as shown in fig. \ref{disp}c), that have both fCE and spinon-like character (see fig. \ref{disp}a, b for a carricature of the excitations). Upon approaching the surface, it possible for the fCE to unbind as a result of reduced $U_{df}/t_d$ in a region of typical size $\sim\xi$, thereby liberating the $d-$electrons and holons, which may Bose condense. The latter leads to confinement and the resulting state is then a decoupled metallic surface. Only the top and bottom surfaces are shown for clarity.}
\label{surf}
\end{figure}

Let us now briefly describe a possible mechanism that allows the insulating bulk hosting a CEFL to coexist with a metallic surface. Previously, it has been argued \cite{Coleman3} that the Kondo-screening can be reduced significantly near the surface leading to  ``Kondo-breakdown", 
in which the $f-$moments decouple from the conduction electrons, giving rise to quasiparticles that are lighter. As a result of surface-reconstruction and screening effects \cite{Vojtasurface}, it is also possible that the ratio $U_{df}/t_d$ is smaller close to the boundaries than in the bulk. The weaker attraction between the holon and the conduction electrons can then lead to an unbinding of the fCE close to the surface, thus liberating the holon and the conduction elecrton within a length scale, $\xi$, from the surface (fig.\ref{surf}). The unbound holons can then Bose condense near the surface, confining the gauge-field, thereby rendering the originally neutral fermions with physical charge. In this way, one may recover metallic quasiparticles at the surface as a result of unbinding of the fCE. Moreover, depending on the details of the fCE dispersion (which may be itself topological) and the odd-parity hybridization, $V_{\alpha\beta}(\k)$, it is possible for the metallic quasiparticles at the surface to realize topologically protected surface-states. We leave a discussion of the detailed quantitative theory for future work.

{\bf Phenomenology of CEFL-} Returning now to a description of the bulk, the low-temperature specific heat is dominated by the fluctuation of the fermion-gauge field system. As a result of the gauge-field fluctuations (see Methods section and the supplementary note 2 for a discussion of the low-energy field theory) the low-$T$ specific heat \cite{holstein,Reizer} is given by,
\beq
C = \gamma  T,~~\tn{where}~ \gamma \sim \ln(1/T).
\eeq
Measurements of specific heat in SmB$_6$ do report a linear in $T$ specific heat \cite{flachbart,SS15,thompson}. Moreover the gapless fCE excitations along the neutral Fermi surface contribute to the NMR spin-lattice relaxation rate, $1/T_1$, in the usual way,
\beq
\frac{1}{T_1T} = \tn{const}.
\eeq 
Measurements on SmB$_6$ support such metallic $1/T_1T$ behavior (V. Mitrovic, personal communication) \cite{Korringa}. 
Note however that as a result of strong spin-orbit effects, the above quantity need not be related to the Knight-shift by Korringa's relation.

The mere presence of a charge-gap in the system does not imply a lack of sub-gap optical conductivity \cite{NgLee07}; the only physical requirement is that the conductivity vanish as $\omega\rightarrow0$. We are interested here in the form of $\tn{Re}[\sigma(\omega)]$ at low, but finite, frequencies. We apply the Ioffe-Larkin rule to the (fCE + holon) system \cite{IL}  (see supplementary note 3 for details) and relate the holon-response to a dielectric constant, $\epsilon_b$. We expect the response of the fCE to be similar to that of a metal at low but non-zero frequencies with $\tn{Re}[\sigma_{\tn{ce}}(\omega)]\gg\omega$ and $\tn{Im}[\sigma_{\tn{ce}}(\omega)]\ll\tn{Re}[\sigma_{\tn{ce}}(\omega)]$. Then,
\beq
\tn{Re}[\sigma(\omega)] = \omega^2 \bigg(\frac{\epsilon_b-1}{4\pi} \bigg)^2 \frac{1}{\tn{Re}[\sigma_{\tn{ce}}(\omega)]},
\eeq
where the fCE conductivity can be expressed in the generalized Drude form $\sigma_{\tn{ce}}(\omega) = \rho/(\Gamma(\omega) - i\omega)$, with $\Gamma(\omega)$ a frequency dependent scattering rate and $\rho$ is defined to be the total optical weight. At low $\omega$,  where $|\Gamma(\omega)|\ll\omega$, the real part of the conductivity can be evaluated as \cite{Rosch05,Rosch06},
\beq
\omega^2~ \tn{Re}[\sigma_{\tn{ce}}(\omega)] \approx \rho ~\tn{Re}[\Gamma(\omega)].
\label{Gammaw}
\eeq
Depending on the mechanism responsible for relaxation of currents, one can then obtain different results for $\Gamma(\omega)$; we discuss the different regimes in the methods section. Recent measurements of optical-conductivity in the THz regime in SmB$_6$ \cite{Armitage16} find appreciable spectral weight below the insulating gap, much larger than any imaginable impurity band contribution. 

After integrating out all the matter-fields, the ground-state energy of the system in the limit of weak fields follows from gauge-invariance,
\beq
u(\b,{\vec{B}}) = u_0 + \frac{(\b-{\vec{B}})^2}{2\mu_b} + \frac{\b^2}{2\mu_{\tn{ce}}} + \frac{\vec{B}^2}{2\mu_v} + u_{\tn{osc}}(\b) + ..,
\label{energy}
\eeq
where $\mu_b,~\mu_{\tn{ce}},~\mu_v$ represent the permeability of the gapped holons, composite-excitons and the background `vacuum' respectively; all of these quantities depend on the UV details of the underlying theory. $u_\tn{osc}(\b)$ is the oscillatory component, relevant for our discussion on quantum-oscillations and the ellipses denote additional higher order terms. In the limit of a small $\vec{B}$, the internal $\b$ can optimize itself in order to minimize the energy;  ignoring the oscillatory component in Eq.\ref{energy}, the optimum value is
\beq
\b = \overline{\b} = \alpha \vec{B},~ \tn{with}~\alpha = \frac{\mu_{\tn{CE}}}{(\mu_{\tn{CE}} + \mu_b)}
\label{bB}
\eeq
an $O(1)$ number that is a priori unknown. In the regime where $\mu_{\tn{CE}}\gg\mu_b$, $\b$ locks almost perfectly to the external $\vec{B}$ (i.e. $\alpha\rightarrow1$). 

The period of the oscillations is then (see methods section) \cite{QO},
\beq
\Delta\bigg(\frac{1}{B}\bigg) = \frac{2\pi }{S_\perp^i}\left(1 + \frac{\mu_b}{\mu_{\tn{CE}}} \right)^{-1} = \frac{2\pi \alpha}{S_\perp^i},
\eeq
where $S_\perp^i$ is the cross-sectional area of the fCE Fermi surface sheet $i$ perpendicular to $\vec{B}$. In the limit where $\alpha\rightarrow1$ (i.e. where $\b\rightarrow\vec{B}$), the period is directly related to the volume of the composite exciton fermi surface, but in general it can be significantly different depending on the value of $\alpha$. Including the effect of impurities broadens the Landau-levels and the oscillation amplitude has an additional Dingle suppression $\sim\exp(-1/\omega_{ci}\uptau_i)$ \cite{DS}, where $\uptau_i$ is the elastic lifetime and $\omega_{ci}$ is the effective cyclotron energy in sheet $i$. 

The low temperature thermal conductivity, $\kappa$, is dominated by the fermionic contribution (i.e. the holon, the gauge-field and the phonon contributions are expected to be small compared to the fCE contribution) and there is no difference between the physical thermal conductivity and the conductivity due to the fCE. Let us first estimate the longitudinal thermal conductivity, $\kappa_{xx}\approx\kappa_{xx}^{\tn{ce}}$, due to the composite-excitons. We assume that the fermionic composite excitons form a state akin to an ordinary metal for thermal transport \cite{PAL10}, such that the longitudinal conductivity is given by
\beq
\kappa_{xx}= \sum_{i=1,2} \frac{k_{\tiny{B}}^2 \uptau_i}{9 m_i} \bigg(\frac{2 m_i \ve_F}{\hbar^2} \bigg)^{3/2} T,
\label{kx}
\eeq 
where $\ve_F$ is the Fermi-energy, $m_i$ represent the masses for the two pockets and we have allowed for different lifetimes, $\tau_i$, for the two pockets. At zero magnetic-fields, all of the experiments on SmB$_6$ find a value of $\kappa_{xx}/T$ that extrapolates to zero as $T\rightarrow0$  \cite{SS15,Li16}. There is no consensus yet on whether $\kappa_{xx}/T$ extrapolates to a finite value in the limit of $T\rightarrow0$ at a finite magnetic field \cite{SSnew,Li16}. The presence of a zero-field $T-$linear specific heat combined with an absence of a finite $T-$linear thermal conductivity suggests the presence of either a small zero-field gap that closes at higher fields, or, the presence of localized states. Within the former scenario, it is plausible that at zero-field and at low temperatures, the CE Fermi surfaces undergo pairing to yield a gapped state with a small insulating gap, that can be significantly smaller than the charge-gap.

We note that experimentally, the thermal conductivity measurements have been carried out at very low temperatures ($<1$ K) while the coefficient of the linear in $T$ specific heat is typically extrapolated from higher temperatures. The opening of a small insulating pairing gap at a temperature $T_p$ corresponds to an actual phase transition in $(3+1)-$dimensions (in the Ising universality class) with an associated divergence in the specific heat. Interestingly, a number of experiments report a strong upturn in the specific heat around $\sim1$ K, which is believed to be inconsistent with the usual Schottky contribution. Within the above scenario, it is plausible that the upturn in the specific heat is associated with the onset of the divergence around $T_p\approx 1$K.

It is also useful to estimate the thermal Hall conductivity, $\kappa_{xy}$. In the weak-field regime, as noted previously, the composite excitons move essentially under the effect of an effective magnetic field $\overline\b$ and are subject to the Lorentz force associated with this field.  However note that the two pockets contribute to the thermal Hall response with opposite signs. We know semi-classically that for each pocket
\beq
\kappa^i_{xy} = (\omega_{c,i}\uptau_i) \kappa^i_{xx},
\eeq
where $\omega_{c,i }= e|\b|/m_i = \alpha e |{\vec{B}}|/m_i$ and $\kappa^i_{xx}$ can be read off from Eq.\ref{kx} above. The total thermal Hall response is the difference of the response for the `particle' and `hole'-like pockets.
Observation of a non-zero thermal Hall effect is a good indicator that the parameter $\alpha$ - which determines the magnitude of orbital effects of the external magnetic field - is not too small.
In SmB$_6$, if the quantum oscillations truly arise from the bulk neutral fermi surface of composite excitons as a result of the mechanism proposed above, then that necessarily implies a finite bulk thermal Hall response.  However, we note that since the system is analogous to a compensated semi-metal, the thermal Hall effect is expected to be vanishingly small at higher fields when $\omega_{c,i} \uptau_i \gtrsim 1$.

Let us finally address the fate of the fCE semi-metal phase as it is doped away from the mixed-valence limit by excess $d-$electrons or holes (e.g. by chemical substitution or by gating thin films). There are two natural outcomes: if the holon remains uncondensed, the $d-$electrons (or holes) can form a `small' fermi surface while the neutral fCE fermi surface continues to be present. This phase is the familiar (mixed-valence) fractionalized Fermi-liquid (FL*) \cite{TS04}.  On the other hand, if the holons condense as a result of doping away from the mixed-valence limit, the CE fermi surfaces become Fermi surfaces of physical electrons (and holes) as a result of confinement. The exact outcome is sensitive to microscopic details and is beyond the scope of our discussion here.

The mechanism responsible for the formation of the fermionic exciton is physically distinct from the one responsible for the conventional bosonic exciton \cite{MottMV}. A few recent theoretical studies have tried to address the origin of the low-energy bulk excitations in SmB$_6$ using a variety of different approaches \cite{CooperExc,Baskaran,ColemanSC}. The CEFL is strikingly distinct  from these previous proposals  - unlike Refs. \onlinecite{Baskaran,ColemanSC} the composite exciton is not a Majorana fermion, and unlike Ref. \onlinecite{CooperExc}, the composite exciton has fermi statistics and forms a fermi surface (see supplementary note 4 for a more detailed comparison).

{\bf Discussion-} We have described a phase of matter with a neutral Fermi surface of composite excitons in a mixed-valent insulator with a charge-gap. A number of properties associated with such a phase resembles the experimental results in bulk SmB$_6$. Future numerical studies of the periodic Anderson model in the insulating regime and in the presence of strong interactions may be able to shed light on questions related to energetics and stability of various phases. We also note that more recent measurements on a mixed valence insulator compound different from SmB$_6$, that displays clear bulk quantum oscillations and has metallic longitudinal thermal conductivity down to the lowest measurable temperatures at zero field, in a clear indication of the formation of a Fermi surface of neutral fermions (L. Li, Y. Matsuda, and T. Shibauchi, personal communication).\\

\begin{center}
{\bf Methods}
\end{center}
{\bf Slave-boson representation-} In order to motivate the rationale behind choosing the prescription in Eq.\ref{sb}, recall that the standard slave-boson representation proceeds as,
\beq
\tilde{f}_{\r\alpha} = h^\dagger_\r \chi_{\r\alpha},
\eeq
where $h_\r$ is a spinless bosonic holon with the constraint $h_\r^\dagger h_\r + \sum_\alpha\chi_{\r\alpha}^\dagger \chi_{\r\alpha} = 1$. Consider now the scenario where the $h-$holons are perturbed away from a Mott-insulating state with $\langle h\rangle =0$ and $\langle h_\r^\dagger h_\r\rangle = 1-x$ (where $x$ represents the density of doped holes away from the $4f^6$ configuration). The two representations are then physically equivalent if we make the transformation $h_\r^\dagger \rightarrow b_\r$ and $\langle b_\r^\dagger b_\r\rangle = x$; for a concrete scenario, consider e.g. the quantum rotor model where $h_\r^\dagger = e^{i\theta_\r}$
  and $n_\r^h$ is the boson density conjugate to $\theta_\r$.

{\bf Fermionic composite exciton hopping-} Consider the limit where there is a clear hierarchy of scales: $U_{ff} \gg U_{df} \gg t_d \gg V$ and where $t_d$ is the nearest neighbor hopping for the $d-$electrons. In this regime, the nearest neighbor hopping amplitude for a single fCE is approximately given by (see supplementary note 1)
\beq
t_{\tn{CE}} \sim t_d \bigg(\frac{t_f}{U_{df}}\bigg),
\label{cehop}
\eeq
where $t_f$ is the effective nearest neighbor $\tilde{f}-$hole hopping. There is, in principle, a very strong on-site repulsion set by $U_{ff}$ for the fCE, as a result of the constraint of no double occupancy for the hard-core holons. However, if the binding is not purely on-site and has some finite extent, the repulsion between the fCE can be renormalized down from the bare $U_{ff}$ and the resulting state can be described within a weakly interacting CEFL.

The fermionic exciton is clearly significantly different from the more conventional bosonic exciton \cite{Mott,MottMV,CooperExc} that has been discussed in the context of semimetals and narrow gap semiconductors. The latter arises in the limit where $U_{df}$ dominates over $U_{ff}$. In contrast, as shown above, the fermionic exciton is expected to arise naturally in the limit where $U_{ff}\gg U_{df}$, which is a more realistic regime for mixed-valent systems. Fermionic composite excitons have also been discussed recently in the context of                                                                                                                                                                                                    multi-component quantum hall states \cite{MB16}.

{\bf Low energy field theory for CEFL-} Let us describe the low-energy effective field theory for the CEFL phase described in the main text \cite{TS04}. The composite exciton, $\psi_{\k\alpha,i}$, with $i=1, 2$ representing the two pockets, is coupled minimally to $a_\mu$ and the non-relativistic $b$ holons at a finite chemical potential, $\mu_b>0$, are coupled minimally to $\D a_\mu = a_\mu - A_\mu$ (see supplementary note 3 for a more complete discussion). 
Let us first discuss the form of the gauge-field propagator, $D_{ij}(i\omega_n,\q) \equiv \langle a_i(i\omega_n,\q) ~a_j(-i\omega_n,-\q)\rangle$ where we choose to work in the Coulomb gauge $\nabla\cdot\a=0$, with $\a$ being purely transverse. As a result of the minimal coupling, integrating out the fCE excitations leads to a Landau-damped form of the propagator,
\beq
D_{ij}(i\omega_n,\q) = \frac{\delta_{ij} - q_iq_j/q^2}{\Xi|\omega_n|/q + \beta q^2},
\label{Dij}
\eeq
where $\Xi,~\beta$ are constants determined by details of the fCE dispersion.  

For the specific non-relativistic form of the theory for the holons, there are no holons in the ground state and the only holon self-energy, $\Sigma_b$, contribution arises as a result of the coupling to the gauge-field and $\Sigma_b(i\omega_n,\q) \sim q^2 \bigg(1 + c |\omega_n| \ln(1/|\omega_n|) + ... \bigg)$ at $T=0$, where $c$ is a constant. The above correction is less important than the bare terms in the holon action and can therefore be ignored. 

Finally, as a result of the coupling to the gauge-field fluctuations, the fermions have a self-energy,
\beq
\tn{Im}~\Sigma_\tn{ce}(\omega) \sim \omega,
\eeq
upto additional logarithmic corrections. 
 
{\bf Alternative route to CEFL-} We demonstrate here an alternate route towards arriving at a description of the bulk CEFL phase from a different starting point. Consider a compensated semi-metal with (physical) $d-$electron and $f-$hole pockets. We are interested in driving the semi-metal into an insulating phase in the presence of strong interactions. The Hamiltonian is given by,
\beq
H_{\tn{csm}} = &&-\sum_{\r\r'} t^d_{\r\r'} d^\dagger_{\r\alpha} d_{\r'\alpha} + \sum_{\r\r'} t^f_{\r\r'} f_{\r\alpha}^\dagger f_{\r'\alpha}  \nonumber\\
&& + \sum_{\r\r'}\epsilon_{\alpha\beta}V_{\r\r'} d^\dagger_{\r\alpha} f_{\r'\beta}^\dagger + H_{\tn{int}},
\eeq
where the hoppings $t^d,~t^f$ are positive and $V$ denotes the hybridization. We will specify the form of $H_{\tn{int}}$ momentarily. 

Consider setting $V=0$ for now and using the slave-rotor formalism to represent the electronic operators as,
\beq
d_{\r\alpha} \equiv e^{i\theta_\r} \psi_{\r\alpha},~~f_{\r\alpha} \equiv e^{-i\theta_\r} \chi_{\r\alpha},
\eeq
where the rotor field, $e^{i\theta_\r}$, carries physical charge and the spinful Fermions $\psi_\alpha,~\chi_\alpha$ are electrically neutral. Let $n_\r$ be the boson density conjugate to the rotor field. Then the gauge-invariant states satisfy the constraint : $n_\r + n^\psi_\r - n^\chi_\r = 0$, where $n^d_\r = n^\psi_\r,~n^f_\r = n^\chi_\r$. Let us then consider the interaction term to be of the form,
\beq
H_{\tn{int}} = U\sum_{\r}(n_r^d - n_r^f)^2 \rightarrow U\sum_\r n_\r^2.
\eeq  
It is then clear that at small $U$ (compared to the hoppings), the rotor fields condense $\langle e^{i\theta_\r}\rangle \neq0$ and we recover the compensated semi-metal phase. At strong $U$, one can drive a `Mott'-transition to a phase where the rotor-fields are gapped $\langle e^{i\theta_\r}\rangle = 0$ (i.e. to an insulator) where the $\psi,~\chi$ fermions can form Fermi surfaces, inherited from the original $d,~f$ Fermi surfaces. This is the CEFL phase. Both phases are stable to having a small $V$.

{\bf Optical conductivity of CEFL-} As introduced in Eq. \ref{Gammaw}, in the regime where $\Gamma(\omega)$ arises primarily due to scattering of the fermions off the gauge-field fluctuations and where the effects of static disorder can be ignored (i.e. the mean-free path, $\ell_{\tn{mf}}$, is long), $\Gamma(\omega)\sim\omega^{5/3}$. In three dimensions, this arises from the fCE self-energy, $\tn{Im}\Sigma_{\tn{ce}}(\omega) \sim \omega$ (upto additional logarithms) and includes two extra powers of $|\q|\sim\omega^{1/3}$. Hence, under these set of assumptions, $\tn{Re}[\sigma(\omega)] \sim \omega^{2.33}$.

On the other hand, in the regime where $\Gamma(\omega)$ still arises due to scattering of the fermions off the gauge-field fluctuations, but the finite $\ell_{\tn{mf}}$ modifies the $|\omega|/q$ form in the gauge-field propagator (Eq.\ref{Dij}) around $q\sim \ell_{\tn{mf}}^{-1}$, $\Gamma(\omega)\sim\omega^2$ and $\tn{Re}[\sigma(\omega)] \sim \omega^2$.

Finally note that the fCE density can couple to the local disorder-potential and $\Gamma$ may be dominated entirely by a frequency independent elastic scattering-rate ($\Gamma_0$); then $\tn{Re}[\sigma_{\tn{ce}}(\omega)]\approx \rho/\Gamma_0$ which leads to $\tn{Re}[\sigma(\omega)] \sim \omega^2$. Similarly, as a result of localization effects, it is possible for $\Sigma_{\tn{ce}}(\omega)$ to vanish much faster than $\omega$ such that $\tn{Re}[\sigma(\omega)] \approx \tn{Re}[\sigma_{\tn{ce}}(\omega)]$, in which case results for strongly disordered metals will apply. 

{\bf Quantum oscillations in CEFL phase-} For small fields the energy in Eq.\ref{energy} can be rewritten as,
\beq
u(\b,{\vec{B}}) &=& u_0'({\vec{B}}) + \frac{(\b-\overline\b)^2}{2\mu_{\tn{eff}}}   + u_{\tn{osc}}(\b),\\
u_0'({\vec{B}})  &=& u_0 +\frac{1}{2}\bigg(\frac{1}{\mu_v} + \frac{1}{\mu_b + \mu_{\tn{CE}}}\bigg) \vec{B}^2,
\eeq
and $\mu_{\tn{eff}}^{-1} = \mu_b^{-1} + \mu_{\tn{CE}}^{-1}$. At zero temperature, the oscillatory component is given by \cite{QO},
\beq
u_{\tn{osc}}(\b) &=& \frac{\chi^i_{\tn{osc}}}{2} \left(\frac{|\b|^5}{2\pi S_\perp^i}\right)^{1/2} f\left(\frac{2\pi S_\perp^i}{|\b|} \right),\\
f(x) &=& \sum_{n=1}^\infty \frac{(-1)^n}{ n^{5/2}} \cos(2\pi nx - \pi/4),
\eeq
where $\chi^i_{\tn{osc}}$ sets the scale for the overall amplitude of the oscillations from the Fermi surface sheet $i$. 

{\bf Data availability-} All relevant data are available from the authors upon reasonable request.

{\bf Acknowledgements-} We thank Suchitra Sebastian for sharing many of their unpublished results and thank her and Olexei Motrunich for many stimulating discussions. We also thank Peter Armitage, Nicholas Laurita, Lu Li, Yuji Matsuda and Vesna Mitrovic for discussions and for sharing their data. D.C. is supported by a postdoctoral fellowship from the Gordon and Betty Moore Foundation, under the EPiQS initiative, Grant GBMF-4303, at MIT. While at MIT, I.S. was supported by the Pappalardo Fellowship. T.S. is supported by a US Department of Energy grant DE-SC0008739, and in part by a Simons Investigator award from the Simons Foundation.

{\bf Author Contributions-} D.C., I.S. and T.S. contributed to the theoretical research described in the paper and the writing of the manuscript.

{\bf Competing financial interests-} The authors declare no competing financial or non-financial interests.

\begin{widetext}
\section*{Supplementary Notes 1}
\label{hopping}
{\bf Fermionic composite exciton hopping:} Consider the situation where there is a single site with configuration $4f^5 5d^1$, surrounded by sites with configuration $4f^6$. As $U_\tn{ff}\rightarrow\infty$, and within slave-boson theory, the single site is identified by $n^{\tilde{f}} = n^b =1$. At large $U_{df} (\ll U_{ff})$, the lowest energy configuration will be the state where the binding has taken place with the formation of the CE. In order for the fCE to hop, both the holon as well as the $d-$electron have to hop to the neighboring sites via virtual processes. Let us now estimate this hopping amplitude within a controlled approximation. For this purpose, consider the following related Hamiltonian,
\beq
\tilde{H} = H_d + H_{df} + H_{\tn{t$_f$-J$_\tn{H}$}} + H_{\tn{hyb}},
\eeq
where $H_d$ and $H_{df}$ represent the Hamiltonian for the $d-$electrons and the repulsive density-density interactions with strength $U_{df}$ between the $d$ and $f-$electrons respectively. We have introduced $H_ {\tn{t$_f$-J$_\tn{H}$}}$ as an effective model for the $\tilde{f}-$hole with,
\beq
H_ {\tn{t$_f$-J$_\tn{H}$}} = P_{\tn{G}} \bigg[-t_f\sum_{\langle \r\r' \rangle}\tilde{f}^\dagger_{\r\sigma}\tilde{f}_{\r'\sigma}  + J_H \sum_{\langle\r\r'\rangle} \S_\r\cdot\S_{\r'}\bigg] P_{\tn{G}},\nonumber\\
\label{htj}
\eeq
where $t_f$ represents the nearest neighbor $\tilde{f}-$hole hopping amplitude and $P_G$ denotes the Gutzwiller projection operator that forbids double occupancy at any given site. Finally, $H_{\tn{hyb}}$ is the term ($\sim V$) responsible for hybridization between the $d$ and $f-$electrons. We begin by setting $V=0$, which results in an enhanced $U_d(1)\times U_f(1)$ symmetry associated with the two conserved fermion numbers. Furthermore, as mentioned earlier, let us restrict ourselves to the regime where $U_\tn{ff} \gg U_{\tn{df}} \gg t_d \gg V$. In this limit and within the slave-boson treatment introduced earlier, the fermionic composite exciton hopping is given by,
\beq
t_{\tn{CE}} \sim \frac{t_d ~t_b}{U_{\tn{df}}}.
\eeq
The holon hopping, $t_b$, can now be estimated using a self-consistent mean-field treatment of the Hamiltonian defined in Eq.\ref{htj} above,
\beq
H_ {\tn{t$_f$-J$_H$}} \rightarrow &&-t_f\sum_{\langle \r\r' \rangle} \langle \chi^\dagger_{\r\alpha} ~\chi_{\r'\alpha}\rangle b_\r^\dagger b_{\r'} \nonumber\\
&&- \sum_{\langle \r\r' \rangle}\bigg(t_f \langle b_\r^\dagger b_{\r'} \rangle + J_H  \langle \chi^\dagger_{\r\alpha}~ \chi_{\r'\alpha}\rangle \bigg) ~\chi_{\r\sigma}^\dagger \chi_{\r'\sigma},\nonumber\\
\eeq
where we have ignored pairing terms for the spinon fields. The effective holon hopping will then simply be given by $t_b \sim t_f$ with an $O(1)$ coefficient, as long as the spinons are in a state (e.g. with a Fermi-surface) where $\langle \chi_\k^\dagger \chi_\k\rangle = n^\chi_\k$ is not flat as a function of $\k$. This leads to the estimate of the fCE hopping in Eq. {\color{blue}17}. In the presence of a small but finite $V$, the resulting fCE and spinon bands hybridize and yield a semi ``metal", as discussed in the main text. From the above equation, it is also possible to read off the self-consistently generated hopping for the spinons, $t_\chi$.

\section*{Supplementary Notes 2}
{\bf Low-energy field theory:}
In the CEFL phase, the low-energy degrees of freedom are the fCE, the spinons and the gapped holons that are all minimally coupled to a dynamical gauge-field. The fCE and the spinons hybridize to yield a compensated semi-metal with `particle-like' and `hole-like' pockets. Let us then write down the resulting low-energy effective field theory. We continue to denote the resulting neutral fermions (which are superpositions of the fCE and the spinon) as $\psi_{\k\alpha,i}$, where we have introduced an additional label $i=1, 2$; $i=1$ for the particle-like pocket and $i=2$ for the hole-like pocket. As described earlier, the $\psi$ fermions couple minimally to $a_\mu$ and the non-relativistic $b$ holons couple minimally to $\D a_\mu = a_\mu - A_\mu$. The Lagrangian is given by
\beq
\L &=& \L_\psi + \L_b + \L_{\psi b} + \L_a,\\ 
\L_\psi &=& \psi_{\alpha,i}^\dagger (\d_\tau - i a_0 - \mu_i) \psi_{\alpha,i}  - \frac{1}{2m_i} \psi_{\alpha,i}^\dagger (-i\nabla - \a)^2 \psi_{\alpha,i},\nonumber\\ \\
\L_b &=& b^*(\d_\tau - i \D a_0 - \mu_b)b - \frac{1}{2m_b} b^* (-i\nabla - \D\a)^2 b \nonumber \\
 &+&  \frac{u}{2} |b|^4 + ... \\
 \L_{\psi b} &=& g_0 ~|b|^2 ~\psi_{\alpha,i}^\dagger \psi_{\alpha,i}  + ...,\\
\L_a &=& \frac{1}{e^2} (\epsilon_{\mu\nu\lambda} \partial_\nu a_\lambda)^2. 
\eeq
The form of the above Lagrangian is similar to theories considered earlier in a different context \cite{TS04}. Note that we have allowed density-density interaction terms between the holon density and composite exciton density in $ \L_{\psi b}$ above; these terms are formally irrelevant close to the transition $\mu_b=0$. $\L_a$ contains a Maxwell term for the emergent gauge-field, but as discussed in the methods section, integrating out the fermions leads to a landau-damped form of the propagator for the transverse-component of the gauge-field, $D_{ij}$, as in Eq.{\color{blue} 18}. The time component of the gauge-field couples to the density and does not lead to any singular non-Fermi liquid behavior. 

In addition to the properties of the holon self-energy, $\Sigma_b$, at $T=0$ that we already described in the methods section, it is important to recall that at a finite temperature $T>0$, $\Sigma_b(0,\vec{0})\sim u T^{3/2}$ and therefore the properties associated with  the holon are determined by an interplay of $\mu_b$ and the above ``thermal" mass. In three-dimensions, there is no phase-transition associated with condensation of $\langle b\rangle$ at $T>0$. 

\section*{Supplementary Notes 3}
{\bf Ioffe-Larkin composition rules:}
We derive the Ioffe-Larkin sum rules \cite{IL} in this supplementary note. From the field theory description introduced in the previous supplementary note, integrating out the vector potential $\a$ leads to the constraint $\bf{j}_\psi + \bf{j}_b = 0$, where the $\bf{j}$ represent the respective currents, i.e. the holons and fermions can only move subject to this constraint. As a result of this constraint, the net electrical response of the system will correspond to the sequential (i.e. `series') circuit of the two individual `resistances'. Suppose we now integrate out the matter fields altogether and obtain an effective action purely in terms of the external and internal gauge fields,
\beq
S_{\tn{eff}}[\A,\a] &=& \frac{1}{2}\int d^3\r~d\tau~\bigg[ \a~\Pi_\psi~\a + (\A-\a)~\Pi_b~(\A-\a) \nonumber\\
&&~~~~~~~~~~~~~~~~~~~+\A~\Pi_0 ~\A + ...\bigg],
\eeq 
where we have restricted ourselves to the simplest quadratic action and $...$ denote additional contributions which may arise e.g. from the singular re-arrangement of the Fermi surfaces in the presence of a finite $\b=\nabla\times\a$. The coefficients $\Pi_\psi$ and $\Pi_b$  denote the full response functions due to the neutral fermions and gapped holons respectively. We have also included $\Pi_0$, that arises from the trivial background for the sake of completeness. The response functions to leading order in small $\omega, \q$ are given by,
\beq
\Pi_\psi(\omega,\q) &=& i\omega~\sigma_\psi(\omega,\q) - \chi_\psi\q^2,\\
\Pi_b(\omega,\q) &=& i\omega~\sigma_b(\omega,\q) - \chi_b\q^2,
\eeq
where $\sigma_{\psi,b}$ and $\chi_{\psi,b}$ denote the conductivities and diamagnetic susceptibilities respectively (Note that $\chi_{\psi,b}\equiv\mu_{\tn{ce},b}^{-1}$, as defined in the main text). Since $\a$ is a dynamical field and the path-integral sums over all allowed configurations of $\a$, we can integrate it out and determine the effective action purely in terms of the external vector potential, $\A$. The resulting action is given by,
\beq
S_{\tn{eff}}[\A] = \frac{1}{2}\int d^3\r~d\tau~ \A\bigg[\Pi_0 + \frac{\Pi_\psi\Pi_b}{\Pi_\psi+\Pi_b}\bigg]\A
\eeq
Let us now first focus on the limit of $\q\rightarrow0$ and finite $\omega$ (i.e. there is a uniform electric field). In this limit, the net conductivity of the system can be read off from the second term above as,
\beq
\sigma(\omega) = \frac{\sigma_\psi(\omega)~\sigma_b(\omega)}{\sigma_\psi(\omega) + \sigma_b(\omega)},
\eeq
which is the promised sequential response of the two resistors. Note that one can arrive at the same formula in terms of the shift of the gauge field $\a$ and its resulting `backflow' effect. In the limit of $\omega\rightarrow0$ (i.e. dc-limit) and at $T=0$, Re$[\sigma_b]=0$ and hence one obtains insulating response. The above expression is also the starting point for the discussion of optical conductivity \cite{NgLee07}, as explained in the main text. 

Let us now consider the other case where $\omega\rightarrow0$ and $\q$ is finite (i.e. there is a uniform magnetic-field). Here, one can read off the net diamagnetic response as,
\beq
\chi = \frac{\chi_\psi ~\chi_b}{\chi_\psi + \chi_b},
\eeq
in an electrical insulator. This is precisely the form of the diamagnetic response that one would obtain starting from the free-energy in Eq. {\color{blue} 11} for $\b=\alpha {\bf B}$ (Eq.{\color{blue} 12}); the resulting locking mechanism is then responsible for Landau quantization of the neutral Fermi-surface and quantum oscillations \cite{QO}. 

\section*{Supplementary Notes 4}
{\bf Comparison to other theoretical proposals for SmB$_6$:}
In this supplementary note, we compare our theoretical proposal for the CEFL with other proposals that have made an attempt to account for at least some of the anomalous features observed in experiments on SmB$_6$. The other proposals can be broadly classified as follows: (i) ``Magnetic-breakdown" in a small-gap insulator \cite{CooperMB,FWMB} - If the typical cyclotron energy, at large enough magnetic fields, is larger than the insulating gap, this can lead to Landau quantization and an analogue of breakdown effects (but as a function of energy). However within this picture, in the low temperature and zero magnetic field limit, the system is still an insulator and does not have any of the thermodynamic or optical signatures associated with the experimental observations described earlier. (ii) ``Majorana" Fermi-surfaces \cite{Baskaran,ColemanSC} - Within this picture, there is a Fermi-surface in the bulk, where the excitations are Majorana-like, instead of being complex fermions of the type we propose (i.e. fCE). In particular in the Majorana-based interpretation, the zero magnetic field state in the limit of zero temperatures is a superconductor, and not an electrical insulator. It is important to note that in an incompressible phase, all excitations carry a well defined charge; the Majorana fermions are objects where the anti-particle is identical to the particle itself and therefore carries no charge. Since there are no neutral fermionic excitations in a system described by an electronic Hilbert space in the UV, these objects can only emerge in the IR and be necessarily non-local. The only known route of doing this theoretically is to couple them to an emergent gauge-field; the non-local Majorana fermions can be coupled to a discrete (e.g. in the simplest case a Z$_2$) gauge field. In three dimensions, there is then necessarily a finite temperature phase transition associated with transition into the phase with deconfined, non-local Majorana excitations. This should be seen e.g. as a divergence in the specific heat at the transition. Within our framework, there is no finite temperature phase transition into the CEFL phase with a deconfined $U(1)$ gauge-field, even in three dimensions. (iii) Gapped (conventional) excitonic insulators \cite{CooperExc} - As a result of a large joint density of states, it is possible that the system is close to a conventional (bosonic) excitonic instability with a finite momentum, $\Q$ (governed by details of the band-structure), but with a small gap, $\Delta$. If the energy gap is small along a ring of momentum around $\Q$ (e.g. like a roton), in the limit of temperatures larger than the gap it can give rise to power-law features in some thermodynamic properties (as opposed to exponential in $\Delta/T$). However, one concern is that this requires some fine-tuning as $\Delta$ needs to be small enough such that the experimentally accessible temperatures are already larger than the gap. At the same time the system needs to avoid undergoing the excitonic instability into a density-wave state (with wavevector $\Q$), since there is no experimental evidence of the system being close to any density-wave instability. 

\end{widetext}
\bibliographystyle{apsrev4-1_custom}
\bibliography{MVI}

\end{document}